\documentclass[aip,graphicx,reprint,onecolumn]{revtex4-1}

\usepackage{graphics}
\usepackage{epsfig}
\usepackage{mathrsfs}
\usepackage{amsbsy}
\usepackage{amssymb}
\usepackage{amsmath}

\usepackage{natbib}

\usepackage[usenames,dvipsnames]{color}

\begin{document}

\title{Fully localised nonlinear energy growth optimals in pipe flow}

\author{Chris C.T. Pringle}
\affiliation{Applied Mathematics Research Centre, Coventry University, Coventry, UK}  
\author{Ashley P. Willis}
\affiliation{School of Mathematics and Statistics, University of Sheffield, Sheffield, S3 7RH UK}
\author{Rich R. Kerswell}
\affiliation{Department of Mathematics, University of Bristol, Bristol, BS8 1TW, UK}

\begin{abstract}
A new, fully-localised, energy growth optimal is found over large
times and in long pipe domains at a given mass flow rate. This optimal
emerges at a threshold disturbance energy below which a nonlinear
version of the known (streamwise-independent) linear optimal (Schmid
\& Henningson 1994) is selected, and appears to remain the optimal up
until the critical energy at which transition is triggered. The form
of this optimal is similar to that found in short pipes (Pringle et
al.\ 2012) albeit now with full localisation in the streamwise
direction. This fully-localised optimal perturbation represents the
best approximation yet of the {\em minimal seed} (the smallest
perturbation capable of triggering a turbulent episode) for `real' (laboratory) pipe flows.


\end{abstract}

\maketitle

\section{Introduction}
In wall-bounded shear flows such as pipe flow, transition to
turbulence remains a problem of great theoretical and practical
importance.  The transition is typically abrupt, occurs at flow rates
for which the underlying base flow is stable, and is triggered by
disturbance amplitudes much smaller that the ensuing turbulent
state. Whether turbulence is triggered or not is also very dependent
on the form of the disturbance\citep{darbyshireM95,faisstE04} and
efforts to identify the best `shape' have revolved around examining
the non-normality of the linearised Navier-Stokes operator around the
base state.\citep{farrell88,farrell89,gustavsson91, butlerF92,
  schmidH92, trefethenSci93, bergstrom93, reddyH93, schmidH94,
  zikanov96, reddy98} This has identified a number of processes by
which a disturbance can grow in energy despite the flow being linearly
stable before it has to ultimately decay.\citep{schmid07}

Recently, the approach of finding optimal perturbations that maximise
growth over a finite time has been extended to retain the nonlinearity
of the Navier-Stokes equations.\citep{PK10, cherubini10,
  monokrousos11} Computationally, this is a very intensive procedure
and so far only small computational domains\citep{PK10, monokrousos11,
  PWK12, rabin12} or short integration
times\citep{cherubini10,cherubini11,cherubini13} have been used to
demonstrate feasibility of the approach. Nonetheless, it serves as a
basis for a proposed procedure for finding the disturbance of the
smallest energy -- the {\em minimal seed} -- which can trigger
transition.\cite{PWK12, KPW14} From a dynamical systems perspective,
this minimal seed is the closest point of approach (in the energy
norm) of the basin boundary of turbulence (or `edge' if the turbulent
state is only transient) to the basic state and it is clearly of
fundamental interest in the problem of transition.

A two stage process has been proposed\citep{PWK12} in which the
maximum energy growth across a given time horizon is sought over all
disturbances of a given initial energy (the `nonlinear transient
growth problem'), and then the initial energy is increased until a
rapid increase in the energy growth is identified. For very large
target times this growth increase tends to a discontinuous jump, and
the first optimal initial condition (as the initial energy increases)
to achieve heightened growth, and thereby be outside the laminar
state's basin of attraction, is then an approximation of the minimal
seed (see the review \cite{KPW14}). 
For more moderate optimisation times the jump in energy growth
is smoothed, and there is a window in initial energy where a new
nonlinear optimal is more efficient than the linear problem's optimal,
but where transition cannot be triggered. Provided the time is still
larger than the transition time, the amplitude of the minimal seed is
also the maximum amplitude of initial perturbation for which
convergence is possible. If much shorter times are chosen, it is
possible to converge at energies well above this critical
amplitude.\cite{cherubini13} Intriguingly, provided a long enough time
is used, these calculations suggest that the minimal seed should be a
fully localised disturbance and therefore of immediate interest to
experimentalists.

The purpose of this letter is to carry out this procedure over a long
pipe domain to demonstrate that the minimal seed is indeed fully
localised when the computational domain is sufficiently long
and to obtain the best estimate yet of what the actual
    minimal seed is for long (real) pipes. The presented work focuses
  on maximising energy growth (the ratio of energy at time $T$ to
  initial energy, $G:=E(T)/E(0)$) for an intermediate choice of
  time. In the 25 diameters ($25D$) long pipe being considered, the
  computational demands of this are already heavy. Nonetheless, the
  results are shown to be close predictors of the true minimal
  seed. The paper is split into five further sections: \ref{sec:maths}
  formulates the problem; \ref{sec:nlop} presents and describes the
  new fully localised nonlinear optimal; \ref{sec:L} explores this
  localisation and shows that the results from much shorter pipes
  (e.g. $5D$) are closely related; \ref{sec:T} examines the importance
  of the choice of optimising time and demonstrates that the observed
  optimal is largely insensitive to the choice of $T$, unless
  exceptionally small values are taken; \ref{sec:disc} contains a
  discussion of the results.

\section{Formulation}\label{sec:maths}

We consider the problem of a Newtonian fluid in a straight pipe of
circular cross-section. A constant mass-flux is imposed along the
pipe, forcing the fluid to flow through it at a constant
rate. Nondimensionalising by the pipe radius ($D/2$) and the mean
axial velocity ($U$), the governing equations of motion are
\begin{equation}
\partial _t \mathbf{u} + \mathscr{U}\partial_z\mathbf{u} +u\mathscr{U}' \mathbf{\hat{z}}
+\mathbf{u}.\nabla \mathbf{u}= -\nabla p +Re^{-1}\nabla^2\mathbf{u}
\end{equation}
where $\mathscr{U} \mathbf{\hat{z}}=2(1-s^2) \mathbf{\hat{z}}$ is the
underlying laminar flow to which $\mathbf{u}=(u,v,w)$ is the not-necessarily-small perturbation 
in cylindrical coordinates
$(s,\phi,z)$. $Re:=UD/\nu$ is the Reynolds number. Periodic boundary conditions are imposed across the ends of the pipe (i.e. in $z$) and no slip conditions on the pipe wall.

We wish to identify the perturbation with initial energy $E_0:=E(0)$
that will undergo the most growth over a given period of time ($T$),
and to this end we employ the usual variational approach \citep{KPW14}. The
functional
$$
\mathscr{L}:=\langle \tfrac12 \mathbf{u}(\mathbf{x},T)^2\rangle -
\lambda \bigg[ \langle \tfrac12\mathbf{u}(\mathbf{x},0)^2\rangle - E_0 \bigg]
-\int_0^T \langle \boldsymbol{\nu} \cdot \bigg[
\partial _t \mathbf{u} + \mathscr{U}\partial_z\mathbf{u} +u\mathscr{U}' 
\mathbf{\hat{z}}+\mathbf{u}.\nabla \mathbf{u}+\nabla p -Re^{-1}\nabla^2\mathbf{u}
\bigg] \rangle \mathrm{d}t
$$
\begin{equation}
\hspace{4cm}-\int_0^T \langle \Pi \nabla \cdot \mathbf{u} \rangle \mathrm{d}t 
-\int_0^T \Gamma \langle \mathbf{u} \cdot \mathbf{\hat{z}} \rangle 
\mathrm{d}t ,
\end{equation}
is maximised numerically in the manner laid out in ref \citep{PWK12}. Throughout this paper we will consider two quantities
\begin{equation}
e(z,t):=\frac12\int_0^{2\pi}\int_0^1 \mathbf{u}^2\,s\,ds\,d\phi \qquad \textrm{and} \qquad E(t):=\frac12\int_0^{2\pi/\alpha}\int_0^{2\pi}\int_0^1 \mathbf{u}^2\,s\,ds\,d\phi\, dz=\int_0^{2\pi/\alpha} e(z,t)\, dz.
\end{equation}
$E(t)$ is the total energy of the perturbation while $e(z,t)$ is the energy
per unit length of the perturbation at a given axial position along
the pipe at a given time. We also consider the roll and streak energies
\begin{equation}
E_{uv}(t):=\frac12\int_0^{2\pi/\alpha}\int_0^{2\pi}\int_0^1 (u,v,0)^2\,s\,ds\,d\phi\, dz \qquad \textrm{and} \qquad E_w(t):=\frac12\int_0^{2\pi/\alpha}\int_0^{2\pi}\int_0^1 (0,0,w)^2\,s\,ds\,d\phi\, dz,
\end{equation}
with the equivalent versions $e_{uv}(z,t)$ and $e_w(z,t)$ defined in the expected way.

All calculations are performed with 64 finite difference points in $s$ (concentrated near the boundary)
and azimuthal Fourier modes running from $-23$ to $23$. For a pipe of
length $L=25D$, we use axial Fourier modes between $-128$ and $128$,
while for other lengths of pipe this is adjusted to keep
the resolution unaltered. Throughout we use $Re=2400$ and, except
where indicated otherwise, we take the optimisation time to be
$T=T_{lin}=29.35\, D/U$ -- the time that maximises the linear optimal growth
at this Reynolds number.

\section{Localised nonlinear optimal}\label{sec:nlop}

\begin{figure}
\begin{center} 
\resizebox{\textwidth}{!}{\includegraphics[angle=0]{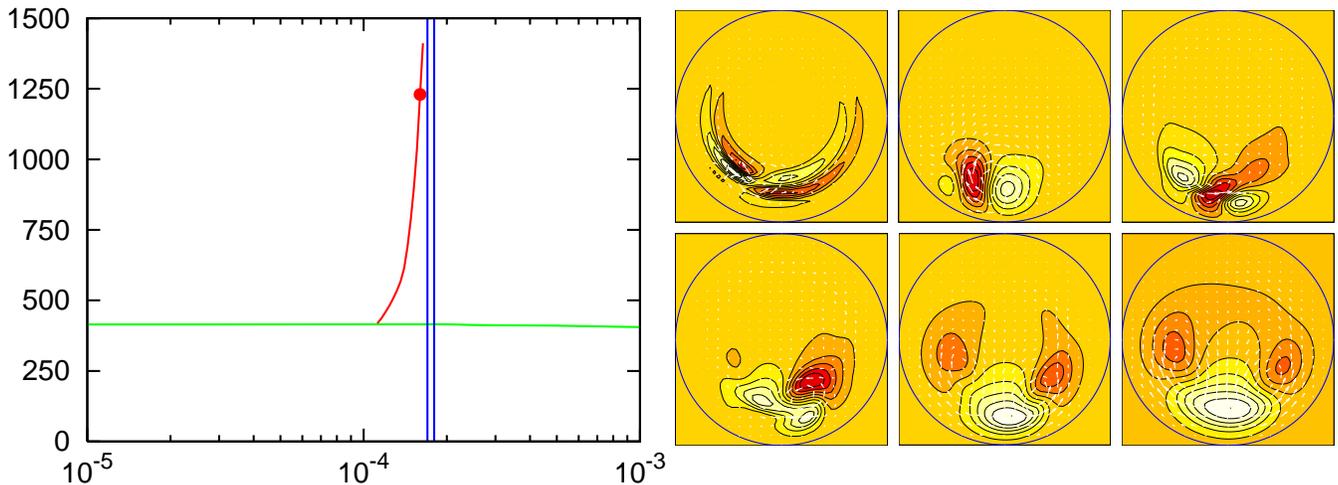}}
\end{center}
\caption{\textbf{Left}: Growth $:=E(T_{lin})/E(0)$ as a function of
  initial energy $E(0)$ for $T=T_{lin}$. The green (almost flat) line
  is the result of a streamwise-independent, nonlinear
  calculation. The optimal (QLOP) observed is very similar to the
  linear optimal. The red (steeply climbing) line shows the new $3D$
  optimal NLOP, while the vertical blue lines represent the interval
  within which turbulent seeds begin to appear. The solid dot
  indicates the nonlinear optimal at $E_0=1.6\times10^{-4}$, used as
  the exemplar localised optimal throughout. \textbf{Right}: The
  evolution of the NLOP in a $25D$ long pipe with
  $E_0=1.6\times10^{-4}$. The slices are taken at $z$ corresponding to
  the maximum value of $e(z,t)$ at times $t=0$, 1, 2.5, 10, 20 and
  $T_{opt}$ (all in $D/U$) (left to right, top to bottom). Streak
  contour levels are varied between slices to show the structure of
  the growing disturbance.}
\label{fig:G_E0}
\end{figure}

In order to find the localised nonlinear optimal, we performed the
transient growth calculation outlined in section \ref{sec:maths} in a
$8\pi \simeq 25$ diameter long pipe. As with shorter domains, for
initial energies below a certain threshold (here
$E=1.12\times10^{-4}$), a streamwise independent minor variation of
the linear optimal (Quasi-Linear Optimal Perturbation, abbreviated to QLOP) is
found. At energies larger than this threshold, a new three dimensional
optimal (NonLinear Optimal Perturbation, or NLOP) emerges
(Fig.\ \ref{fig:G_E0}, left). The growth produced by the new NLOP
quickly dwarfs the energy growth of the corresponding QLOP as the
initial energy is increased further, until convergence ceases to be
possible and we begin to find turbulent seeds --- perturbations that
lead to a turbulent end state by $t=T$. To determine the precise point
at which convergence fails is beyond the resources available, but it
is bounded as follows $1.7\times10^{-4}<E_{fail}<1.8\times10^{-4}$.

%
%

\begin{figure}
\begin{center} 
\resizebox{0.9\textwidth}{!}{\includegraphics[angle=0,clip=true]{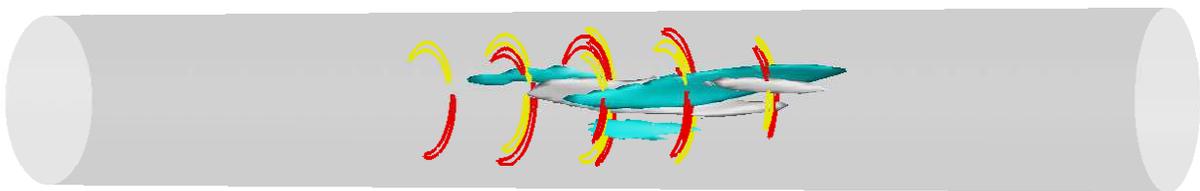}}
\end{center}
\caption{A $7D$ section of the NLOP from a $25D$ pipe for
  $E_0=1.6\times10^{-4}$. The white (cyan) surface is an isosurface of
  where the vorticity is $30\%$ ($-30\%$) of the maximum vorticity in
  the pipe. The yellow (red) lines are contours on cross-sectional
  surfaces of positive (negative) streamwise velocity.  }
\label{fig:iso}
\end{figure}

For the nonlinear optimal corresponding to $E_0=1.6\times10^{4}$, we
plot cross-sections of the perturbation during its development
(Fig.\ \ref{fig:G_E0}, right). The sequence shown is very similar to
that observed in shorter pipes.\citep{PK10, PWK12} Like the previously
known optimals, the initial disturbance is strongly localised in the
cross-sectional plane and unpacks through a complicated
procedure\citep{PWK12} to produce two larger rolls straddling three
streaks.  Unlike those previously reported, however, the optimal found
here is also strongly localised in the streamwise direction with
$99\%$ of the energy contained within a $7D$ section of the pipe --
shown in Fig.\ \ref{fig:iso}. The rolls shown as isosurfaces weave
their way along one side of the pipe, threading through the streak
contours shown at discrete cross-sections along the pipe. These
structures are tightly layered and inclined back into the oncoming
flow. This mirrors the structure of the optimal in shorter pipes
\citep{PWK12} where the initial growth is driven by the Orr-mechanism
in which the layers are tilted up into the underlying shear.

%
%

\begin{figure}
\begin{center} 
\resizebox{0.9\textwidth}{!}{\includegraphics[angle=0]{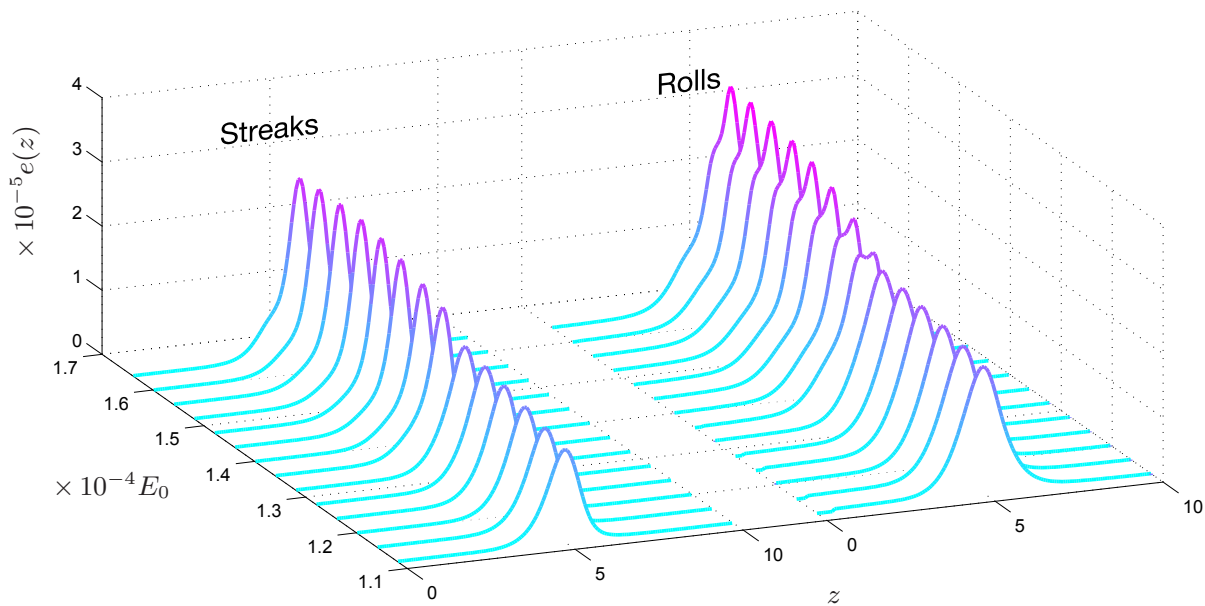}}
\end{center}
\caption{The axial distribution of the initial energy in the streaks ($e_w(z,0)$)
  and the rolls ($e_{uv}(z,0)$) of NLOP in a $25D$ pipe, and how it changes
  as a function of $E_0$ . $z$ is in units of $D$ so only a $10D$ section of the pipe is
  represented. The QLOP on this plot would be represented as a flat
  line of amplitude $E_0/L$.}
\label{fig:roll_strks}
\end{figure}

Localisation is present throughout the energy window in which
convergence to a nonlinear optimal is possible
(Fig.\ \ref{fig:roll_strks}). As the initial energy is varied, the
streak structure remains essentially unchanged.  The roll structure
separates slightly in the axial direction as the initial energy is
increased leading to two slightly distinct peaks for higher energies.

%
%

\section{Effect of L}\label{sec:L}

\begin{figure}
\begin{center} 
\resizebox{0.8\textwidth}{!}{\includegraphics[angle=0]{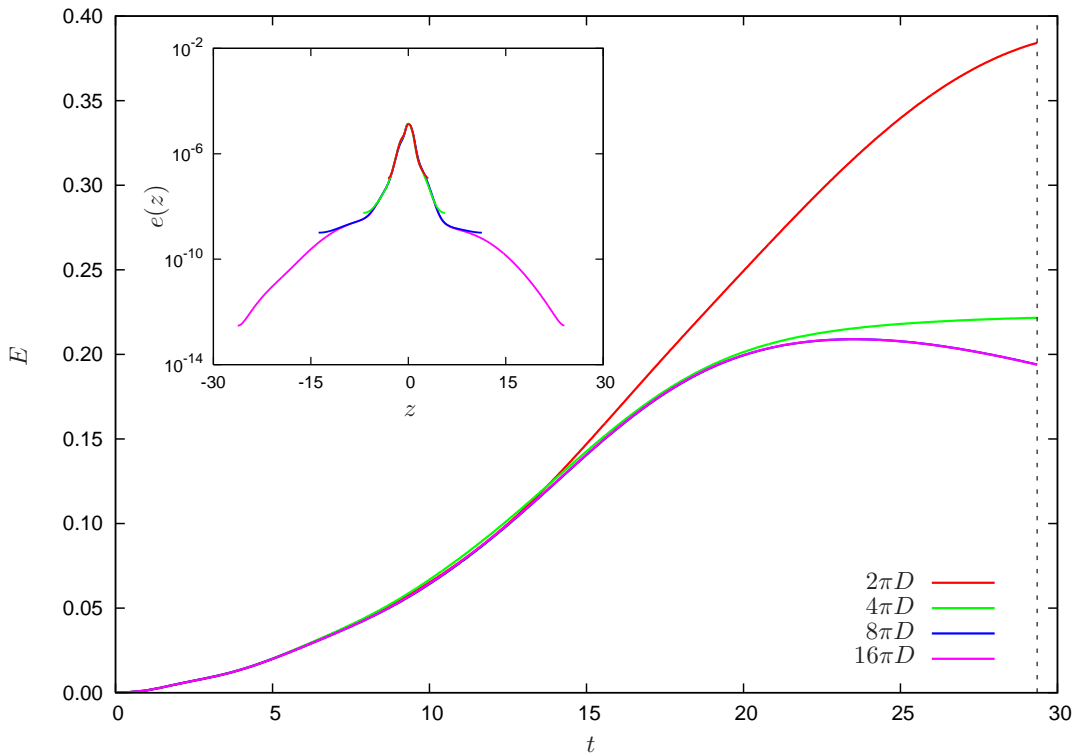}}
\end{center}
\caption{\textbf{Outer:} Evolutions of the NLOPs found in different
  length pipes at $E_0=1.6\times10^{-4}$. All four optimals initially
  have indistinguishable evolutions, before they begin to
  separate. For the two longest domains they remain inseparable
  throughout the evolution period. \textbf{Inner:} The axial
  distribution of the initial energy of the same NLOPs. The central
  structure of the differing optimals closely match, diverging only as
  the ends of the periodic domain are reached.}
\label{fig:L}
\end{figure}

In order to capture a localised optimal, we must make sure that not
only is the optimal initially localised, but that it remains localised
throughout its evolution. The perturbation is expected to swell as it
grows in energy, and the pipe must be long enough that
  as it expands it does not begin to interact with itself through the
  periodic boundary conditions. The degree of
      self-interaction was tested by repeating the nonlinear
  transient growth calculation for pipes of length $L=$
  $2\pi D$, $4\pi D$ and $16\pi D\simeq50D$ at
  $E_0=1.6\times10^{-4}$. In Fig.\ \ref{fig:L}, we plot the energy
  evolution of these optimals along with that of the
  benchmark $8\pi D\simeq 25D$ optimal. In all four
  cases the initial evolution is indistinguishable, but as the
  optimals begin to unfurl along the pipe the evolutions begin to
  diverge. Unsurprisingly this self-interaction has the greatest
  effect upon the optimal in the shortest periodic domain. For the
  $4\pi D$ optimal it is only the very final part of the evolution
  which is affected, while the $25D$ and $50D$ optimals are
  indistinguishable.  The inset of Fig.\ \ref{fig:L} shows the initial
  axial distribution of the energy within these optimals. The central
  portion of each of these optimals align very closely with the
  distributions only diverging as each optimal reaches the ends of its
  periodic domain.

Taken altogether, these results suggest that calculations performed in much
smaller domains are not only able to capture the same mechanisms and
qualitative results as those observed in domains large enough to
capture localised dynamics, but that the optimals found are in fact
precisely the same. The only apparent difference is that for longer
domains there is more space for the energy levels to drop off. This
energy drop off is passive and does not directly influence the form of
the perturbation.

\section{Effect of T}\label{sec:T}

The nonlinear transient growth calculation depends upon the target
optimisation time. It has previously been reported that it is possible
to converge at high initial amplitudes for which turbulence can be
triggered, but this is only if short times are
considered.\citep{PWK12, cherubini13} The amount of growth also
(trivially) varies with this. What is not clear, however, is whether
the \emph{form} of the optimal found in the nonlinear calculation also
depends upon the choice of target time.

\begin{figure}
\begin{center} 
\resizebox{0.8\textwidth}{!}{\includegraphics[angle=0]{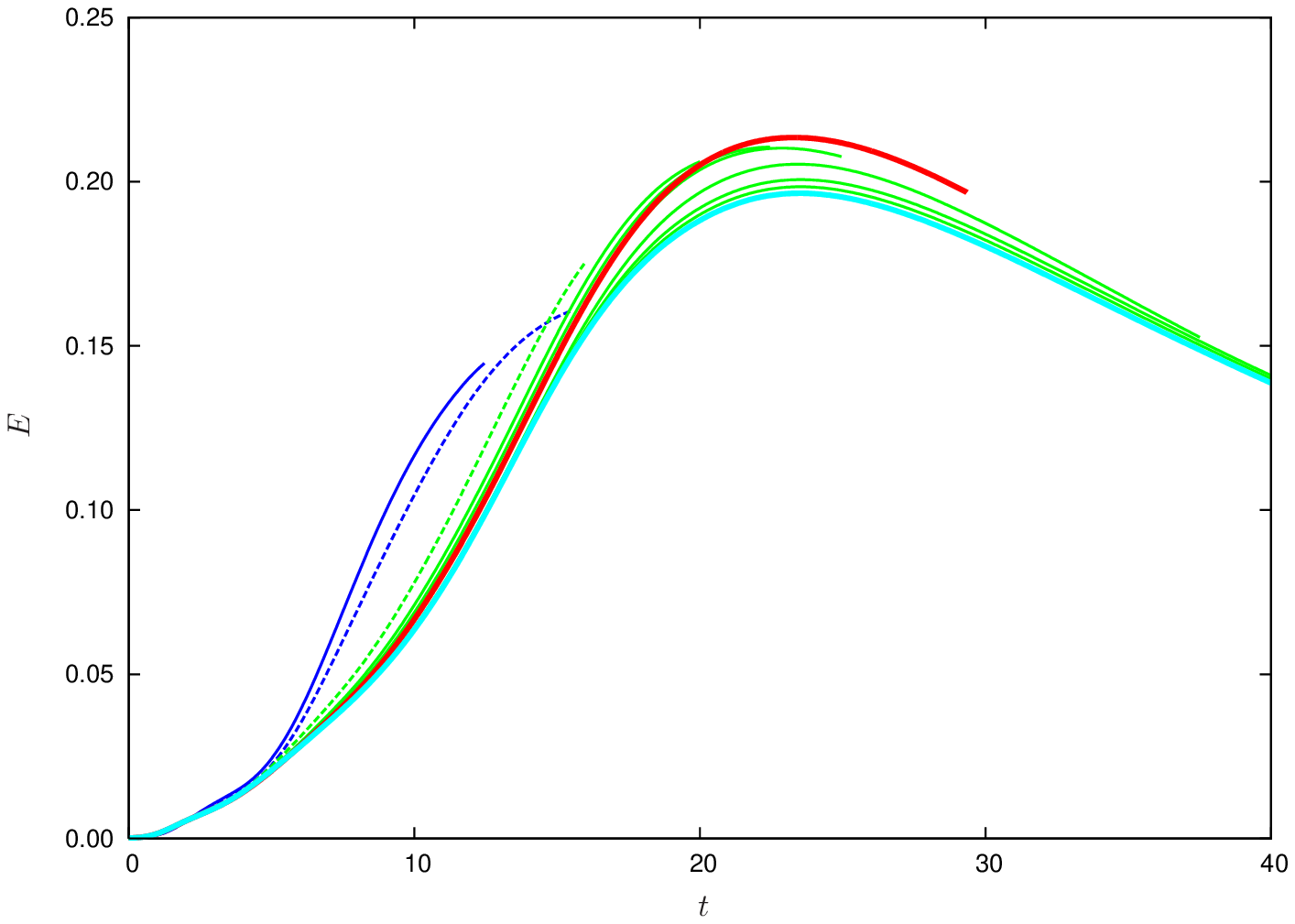}}
\end{center}
\caption{Evolution of optimals with differing $T$. The two blue (dark
  thin )lines have very short optimsation times ($T=12.5 \, D/U$ and
  $15.5\, D/U$).  These two optimals evolve in a notably different
  manner to all the other optimals ($T\geq16 \,D/U$) being
  considered. The optimal corresponding to $T=T_{lin}$ is shown in
  (thick) red, while the largest optimisation time considered
  ($T=100D/U$) is shown in cyan (only the first $40\,D/U$
      is plotted). The two dashed lines correspond to $T=15.5\,D/U$ (blue/dark)
  and $T=16\,D/U$ (green/light) which bracket the abrupt change in evolution of their
  respective optimals.}
\label{fig:T}
\end{figure}

To this end, we performed the nonlinear transient growth calculation
for a range of target times. The evolutions of the optimals found
through this are shown in Fig. \ref{fig:T}. Two separate forms of
evolution are observed. For values of $T\gtrsim16D/U$ the optimals all
evolved in similar manners, reaching a peak energy level at
$T\simeq20D/U$ before decaying away. Smaller values of $T$ give an
optimal which undergoes an accelerated evolution. The eventual maximum
energy levels obtained (were the perturbation allowed to evolve
indefinitely) are lower than before, and the transition between these
two types of evolution is abrupt -- the optimals for $T=15.5D/U$ and
$T=16D/U$ are indicated by dashed lines in Fig.\ \ref{fig:T}.

Unless short times are taken, the optimal observed is relatively
insensitive to the value of $T$ chosen.  Previous
work\citep{PWK12,KPW14} conjectured, and provided evidence that, if
large optimisation times are used then the nonlinear transient growth
algorithm can be used to identify both the minimum amplitude of
disturbance required to trigger turbulence, and the minimum seed that
this equates to. Due to the computationally demanding nature of the
computation even in short domains, in this work we have considered
intermediate optimisation times. Nonetheless, it appears clear that
this is sufficient to observe the \emph{form} of the minimal seed.

%
%

%
%
\begin{figure}
\begin{center} 
\resizebox{0.65\textwidth}{!}{\includegraphics[angle=0]{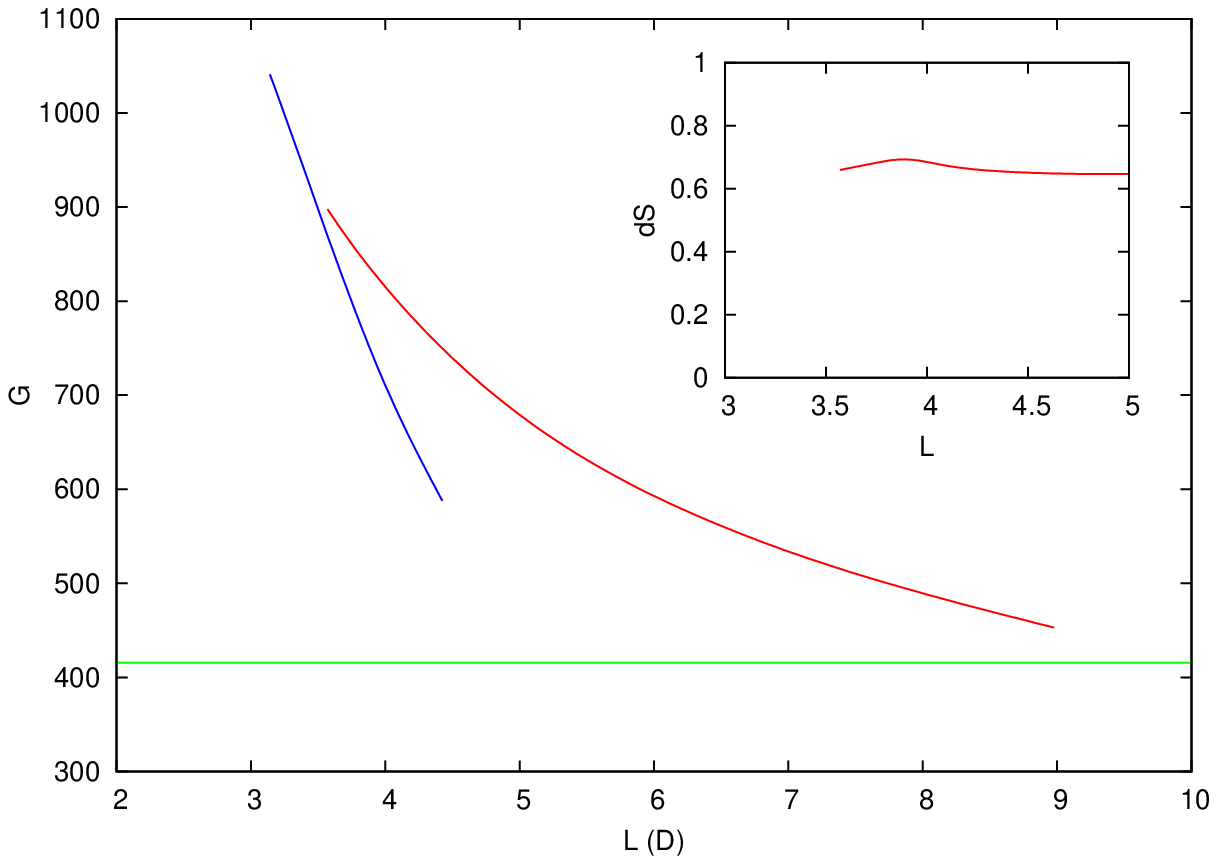}}
\end{center}
\caption{Growth against length of pipe $L$ (in $D$) for the streamwise independent 
      optimal, QLOP (green), the nonlinear optimal, NLOP (red) and the nonlinear optimal
      with shift-and-reflect symmetry enforced with in the domain (blue). The switch 
      in optimal type can clearly be seen at $L\simeq3.5\,D$ with the shift-and-reflect 
      optimal being the global optimal for shorter domains than this. The inset demonstrates that the
      switchover is not a bifurcation but merely a crossing over of
      distinct maximums as the energy in the symmetry-breaking part of
      the large $L$ optimal does not vanish at the switchover
      length. All results are for fixed $E_0=0.8\times10^{-4}$.
      }
\label{switchover}
\end{figure}

\section{Discussion}\label{sec:disc}

We have found the first energy growth optimal, which despite the
underlying equations supporting strictly periodic domain-filling
flows, is fully localised. Reassuringly, this optimal fits well with
previous results. The central structure of the optimal is strikingly
similar to the optimal found in much shorter domains, and appears to
be essentially the same optimal but with an extended region of
exponential decay.  From this it seems clear the energy growth
mechanisms in the localised optimal are essentially the same as for
the optimal found in short pipes\citep{PWK12}, though now the optimal
also expands along the domain as it grows in amplitude. Less clear is
what sets the rate of energy drop off in $z$ -- a strong scaling
appears to be present, but it is not the result of a simple energy
balance.

The importance of $T$ has also been illuminated. In order to estimate
the minimum amplitude of the edge ($E_{c}$) and the minimal seed to
high accuracy, large optimisation target times are required. Despite
this, the form of the minimal seed is accurately revealed by more
intermediate choices. Further, with this choice of $T$, we were able
to find reasonable energy bounds,
$1.7\times10^{-4}<E_{c}<1.8\times10^{-4}$. The upper bound is firm as it comes
  from finding turbulent seeds at this energy.The lower
      bound is less definite from this calculation alone but is
      confirmed by performing larger $T$ calculations, for which $E_c=E_{fail}$
    - see \citep{PWK12,KPW14}. This estimate
  for $E_c$ is consistent with the $5D$ result found much more
  precisely\citep{PWK12} as the slightly lower (since the perturbation 
  can self-interact) value of $E_{c}=1.5\times10^{-4}$.

It is only when we consider much lower choices of $T$ or very short
$L$ that any substantial differences appear. In the case of reducing
$T$, there is an abrupt change where a new optimal emerges which
prioritises fast unsustainable growth. 
A similar change takes place for very short choices of $L$,
where the optimal switches to one exhibiting the shift-and-reflect
symmetry
\begin{equation}
{\cal S}:(u,v,w)(s,\phi,z) \rightarrow (u,-v,w)(s,-\phi,z+L/2);
\end{equation}
see Fig. \ref{switchover} which shows that the switchover between
 the optimals is not a bifurcation as the symmetry-breaking measure
\begin{equation}
dS:= \sqrt{\frac{{\rm Energy \, of \, } (u-{\cal S}u)}{{\rm Energy \,of\,} u}}
\end{equation}
does not vanish there.   This symmetry is unsupportable by a
localised disturbance indicating that this is again a fundamentally
different perturbation. Outside of these two extreme cases, we have
shown that using more computationally viable parameter regimes than
those previously stipulated\citep{PWK12} still allows us to ascertain
insight into the minimal seed and the corresponding critical minimum
amplitude of turbulence. One immediate observation is
    that the minimal seed has 99\% of its energy concentrated in just
    $7\,D$ of the pipe length at $Re=2400$.  This resonates with the
    observation in experimental work \cite{hof03,peixinho07} that once
    disturbances generated by jets become more that $\approx 6\,D$
    long, the ensuing dynamics is largely independent of the
    disturbance length. Recently discovered, localised relative
    periodic orbits in pipe flow also share this lengthscale of 5-10
    $\, D$ as do turbulent puffs: see figure 2 of ref \citep{avila13} and
    figure 4 of ref \citep{chantry2014}.

%
%

In terms of future work, the way is now clear to map out the threshold
energy $E_c$ for transition as a function of $Re$ just as has been
recently done in small-box plane Couette flow \citep{duguet13}
(the imposed streamwise periodicity in \citep{duguet13} is equivalent to
$L=2 \pi D$ here). Our results indicate that using
small-to-intermediate periodic domains (at least in the streamwise
direction) can still yield useful results.

Experimentally, of course, only a small subset $\Sigma$ of all
possible disturbances considered theoretically can actually be
generated. To move the theory closer to this reality just requires
that the optimisation be performed over $\Sigma$ which means simply
projecting the variational derivative of the energy growth with
respect to the initial perturbation down onto $\Sigma$. The greater
theoretical challenge is actually to accurately model the disturbances
routinely generated in the laboratory by injecting or removing fluid
through small holes.\citep{hof03,peixinho07} Adding an artificial
body force temporarily to the Navier-Stokes equations, however, seems
to work well\citep{mellibovsky09}.

Another direction to take this work is into control.  Here the aim
could be to increase $E_c$ by manipulating some aspect of the flow. A
first promising step along these lines has already been made in plane
Couette flow by oscillating the boundaries in their plane and
perpendicular to the shearing direction. \citep{rabin14}

\begin{acknowledgements}
The calculations in this paper were carried out at the Advanced Computing 
Research Centre, University of Bristol.
\end{acknowledgements}
\bibliography{pipeBib}
\end{document}